\documentclass[a4paper, 11pt]{article}

\usepackage{a4wide}
\usepackage{latexsym}
\usepackage{graphicx}
\usepackage{color}
\usepackage{hyperref}
\usepackage{verbatim}
\usepackage{amsmath}

\title{A stochastic evolutionary model generating a mixture of exponential distributions}

\author{Trevor Fenner, Mark Levene, and George Loizou \\
Department of Computer Science and Information Systems \\
Birkbeck, University of London \\
London WC1E 7HX, U.K. \\ \{mark,trevor,george\}@dcs.bbk.ac.uk}

\date{}

\begin{document}

\maketitle

\begin{abstract}

Recent interest in human dynamics has stimulated the investigation of the stochastic processes that explain human behaviour in various contexts, such as mobile phone networks and social media.
In this paper, we extend the stochastic urn-based model proposed in \cite{FENN15} so that it can generate mixture models,
in particular, a mixture of exponential distributions.
The model is designed to capture the dynamics of survival analysis, traditionally employed in clinical trials, reliability analysis in engineering, and more recently in the analysis of large data sets recording human dynamics. The mixture modelling approach, which is relatively simple and well understood, is very effective in capturing heterogeneity in data.
We provide empirical evidence for the validity of the model, using a data set of popular search engine queries collected over a period of 114 months. We show that the survival function of these queries is closely matched by the  exponential mixture solution for our model.

\end{abstract}


\section{Introduction}

Recent interest in complex systems, such as social networks, the world-wide-web, email networks and mobile phone networks \cite{BARA07}, has led researchers to investigate the processes that could explain the dynamics of human behaviour within these networks. Barab\'asi \cite{BARA05} has suggested that the bursty nature of human behaviour, for example, when measuring the inter-event response time of email communication, is a result of a decision-based queuing process. In particular, humans tend to prioritise actions, for example, when deciding which email to respond to, and therefore a priority queue model was proposed in \cite{BARA05}, leading to a heavy-tailed power-law distribution of inter-event times.
The availability of large data sets, such as mobile phone records, has widened the applicability of human dynamics investigation, for example, in an attempt by Schneider et al. \cite{SCHN13} to uncover the characteristics of daily mobility patterns. Human dynamics is not limited to the study of behaviour within communication networks, as can be seen, for example, by a recent proposal of Mitnitski et al. \cite{MITN13}, who apply a simple stochastic queueing model to the complex phenomenon of ageing, in order to illustrate how health deficits accumulate with age.
Moreover, {\em sociophysics} \cite{GALA08,CAST09,SEN14} has similar aims and, in particular, uses concepts and methods from statistical physics to investigate social phenomena and political behaviour.
This methodology often involves the empirical investigation of collective choice dynamics, resulting in the formation of statistical laws governing, for example, how the popularity of movies emerges from social behaviour. In this context, we have recently modelled the survival of search engine queries in the top-10 list of the most popular queries that users have submitted, as an application of human dynamics \cite{FENN15}; see \cite{TATA14} for attempts to predict the future popularity of online content.

\medskip

Survival analysis \cite{LAWL03,KLEI12} provides statistical methods to estimate the time until an event occurs, known as the {\em survival} or {\em failure} time. Typically, an event in a survival model is referred to as a failure, since it often has negative connotations, such as mortality or the contraction of a disease, although it could, in principle, also be positive, such as the time to return to work or to recover from a disease. In the context of email communication mentioned above, an event might be a reply to an email; in the context of popular queries, an event might be dropping out of the top-10 most popular queries.
Traditional applications of survival analysis are in clinical trials \cite{FLEM00}, and in understanding the mechanisms in biological ageing \cite{GAVR01}. The methods used in survival analysis overlap with those used in engineering for reliability life data analysis \cite{FINK08,OCON12}. Reliability analysis in engineering has many applications, for example, in manufacturing processes, and in software design and testing.
However, one can envisage that survival analysis would find application in newer human dynamics scenarios in complex systems, such as those arising in social and communication networks \cite{BARA05,CAND08,MATH13}.

\smallskip

We have been particularly interested in formulating {\em generative models} in the form of stochastic processes by which complex systems evolve and give rise to power laws or other distributions \cite{FENN05,FENN12,FENN15}.
This type of research builds on the early work of Simon \cite{SIMO55}, and the more recent work of Barab\'asi's group \cite{ALBE01} and other researchers \cite{BORN07a}. In the context of human dynamics, the priority queue model mentioned above \cite{BARA05} is a generative model characterised by a heavy-tailed distribution.

In the bigger picture, one can view the goal of such research as being similar to that of {\em social mechanisms} \cite{HEDS98}, which looks into the processes, or mechanisms, that can explain observed social phenomena. Using an example given in \cite{SCHE98a}, the growth in the sales of a book can be explained by the well-known logistic growth model \cite{TSOU02}, and more recently we have shown that the process of conference registration with an early bird deadline can be modelled by bi-logistic growth \cite{FENN13}.
Such research can also be grounded within the field of sociophysics, mentioned above, where critical phenomena are important in demonstrating how the transition to global behaviour can emerge from the interactions of many individuals.
The individuals may be neurons, as in \cite{LOVE12}, where criticality emerges as neuron cooperation,
or people's decisions, as in \cite{PAN10}, where the popularity of movies emerges as collective choice behaviour.

\smallskip

In this work we build on the generative model proposed in \cite{FENN15}, which captures the essential dynamics of survival analysis applications. In the urn-based model presented in \cite{FENN15}, the {\em actors} are called {\em balls}, and a ball being present in $urn_i$, the $i$th urn, indicates that the actor represented by the ball has so far survived for $i$ time steps. An actor could, for example, be a subject in a clinical trial, an email that has not yet been replied to, or an ongoing phone call. As a simplification, we assume that time is discrete and that, at any given time, a single ball may join the system with a fixed {\em birth} probability. In addition, existing balls in the system may be discarded, i.e. one or more {\em mortality} (or death) events may occur, according to some specified {\em mortality function}, which may take a general form. It is evident that, at any given time $t$, we may have at most one ball in $urn_i$, for all $i \le t$.

\smallskip

The probability that, after $t$ steps of the stochastic process outlined above, there is a surviving ball in $urn_i$ is known as the {\em survival function} \cite{KLEI12}. In the context of a medical study, this could represent the probability that a patient survives for longer than a specified time.
In \cite{FENN15} we have shown how the survival function depends on the mortality function.
Two important cases are: (1) when the mortality function is constant then the survival function follows an exponential distribution, and (2) when the mortality function follows a power-law distribution then the survival function follows a Weibull distribution \cite{RINN09}; the latter is often applied to the modelling of {\em life data} (also known as survival data, time to event data, or time to failure data) \cite{KLEI12,OCON12}.

It is interesting to observe that the distribution of the survival function has two parameters, $i$ and $t$, cf. \cite{FENN12}, whereas most previously studied generative stochastic models \cite{ALBE01}, including those in our previous work, for example \cite{FENN05}, result in steady state distributions that are asymptotic in $t$ to a distribution with a single parameter $i$.

\smallskip

Exponential mixture models \cite{MCLA00} have been exploited in survival analysis in a variety of applications,
as mentioned in Section~\ref{sec:mixture} below.
Moreover, it has been shown that the survival function of an exponential mixture model has a decreasing failure rate
\cite{BLOC97,FINK09}, which was first observed in \cite{PROS63} in the context of air-conditioning systems for a fleet of airplanes. To generate exponential mixtures, in Section~\ref{sec:urn}, we extend the urn-based stochastic process of \cite{FENN15} by adding a first step of tossing a biased coin to decide which of two processes becomes active; although the extension is defined for a mixture of two exponentials, it extends naturally to a mixture of several exponentials.
We provide empirical evidence for the validity of the model by analysing the longitudinal data set of popular search engine queries over 114 months that we used in \cite{FENN15}. We show that the survival function of these queries closely follows the distribution generated by our model.

\medskip

The rest of the paper is organised as follows.
In Section~\ref{sec:mixture}, we review mixture models in the context of survival data analysis.
In Section~\ref{sec:urn}, we present our stochastic urn-based model that provides us with a mechanism to model the mixture of exponential distributions. In Section~\ref{sec:app}, we apply the model to the survival of popular search engine queries posted on Google Trends (\url{www.google.com/trends/hottrends}). Finally, in Section~\ref{sec:conc}, we give our concluding remarks.

\section{Mixture models for Failure Time Analysis}
\label{sec:mixture}

Lifetime analysis, also referred to as survival time analysis or failure time analysis \cite{LAWL03}, has a long history of exploiting mixture models \cite{MCLA00}, in particular exponential mixture models, in a variety of applications.
Examples of areas where exponential mixtures have been employed for lifetime analysis modelling are:
the reliability of system components \cite{PROS63}, network performance \cite{FELD98},
actuarial losses \cite{KEAT99}, reliability engineering \cite{FINK09}, and the likelihood of volcanic eruption \cite{MEND09}.

\smallskip

The application we consider in Section~\ref{sec:app}, in the context of survival analysis, is that of predicting the duration of the popularity of search engine queries. We note that this is related to the problem of predicting the popularity of movies \cite{PAN10} or online content \cite{TATA14}. Mixture models are a natural tool for modelling such applications as they are very effective in capturing heterogeneity. In particular, when considering the popularity of queries, a mixture model can differentiate between two types of queries: ones that are very popular for a short period of time and others whose popularity lasts much longer; such an observation has already been made in the context of the popularity of movies \cite{PAN10}.

\smallskip

The {\em survival function} $S(\theta)$ represents the probability that an actor survives for longer than a specified time $\theta$.
Here we restrict ourselves to a mixture of two exponential survival functions, noting that this can readily be extended to a mixture of more than two such functions. The mixture rates of the distribution, $p_1$ and $p_2$, sum to one, and the {\em mortality rates}, $\lambda_1$ and $\lambda_2$, of the mixture model (also known as the {\em failure rates}) are constants. These define a {\em discrete hazard function} \cite{BRAC03}, i.e. the conditional probability that a failure occurs at a specified time instant given that no failure occurred previously. We therefore define the mixture survival function $S(\theta)$ as
\begin{equation}\label{eq:exp-mix}
S(\theta) = p_1 \exp(- \lambda_1 \theta) + p_2 \exp(- \lambda_2 \theta),
\end{equation}
where $p_1+p_2=1$.

\smallskip

The mortality rate of the mixture model \cite{FINK09} is now given by
\begin{equation}\label{eq:lambda-mix}
\lambda(\theta) = \pi_1(\theta) \lambda_1 + \pi_2(\theta) \lambda_2,
\end{equation}
where the time-dependent probabilities $\pi_k$ are defined as
\begin{equation}\label{eq:pi-mix}
\pi_k(\theta) = \frac{p_k \exp(-\lambda_k \theta)}{p_1 \exp(-\lambda_1 \theta) + p_2 \exp(-\lambda_2 \theta)}.
\end{equation}

In \cite{BLOC97,FINK09} it was shown that the mortality rate $\lambda(\theta)$ of the mixture model is decreasing in $\theta$.
Moreover, if we assume that $\lambda_1 \le \lambda_2$, then $\lambda(\theta) / \lambda_1$ converges to $1$ as $\theta$ tends to infinity; this is referred to in the literature as a {\em decreasing failure rate}.

\smallskip

For comparison purposes, we will also make use of the Weibull survival function considered in \cite{FENN15}, with shape parameter $\rho$ and scale parameter $\gamma$, which is given by
\begin{equation}\label{eq:weibull}
\exp\left(- \gamma \ \theta^{1+\rho} \right).
\end{equation}

It is important to note that, when the shape parameter satisfies $-1 < \rho < 0$, then it can be expressed as a mixture of exponentials, since it is a {\em completely monotone function} \cite{JEWE82,FELD98};
see \cite{FELL71} for the underlying theory.

\section{An Evolutionary Urn Transfer Model}
\label{sec:urn}

We next formalise a stochastic urn model based on the survival model for human dynamics introduced in \cite{FENN15}, in order to generate the mixture of two exponential distributions. This is done by first generating a mixture distribution by tossing a coin with mixture probabilities $p_1$ and $p_2$, where $p_1 + p_2 = 1$, to decide which process is active, and then employing a constant mortality rate parameter $\lambda_k$ in order to generate an exponential distribution, as in \cite{FENN15}. We note that this mechanism extends naturally to a mixture of more than two exponentials by deploying a multi-sided ``coin''.

\smallskip

We assume a countable number of urns, $urn_0, urn_1, \ldots \ $, where a {\em ball} (or {\em actor}) in $urn_i$ is said to be of {\em age} $i$. Initially, all the urns are empty. At the start of the stochastic process, we toss the coin, as described above, to decide which of two processes will be active. From then on, at any time $t$, $t \ge 0$, a new ball may be {\em born},
with probability $b$, and any such new ball is inserted into $urn_0$; in addition,
an existing ball of age $i$ can {\em die} by being discarded from $urn_i$, for all $i > 0$.

\smallskip

For a given age $i$ and time $t$, we let the  {\em mortality rate} $\mu_k(i,t)$ be the probability that a ball in $urn_i$ dies at time $t$, where $k=1$ or $2$ depending on which process is active. We always require that $\mu_k(0,t)=0$ for all $t$, and that $\mu_k(i,t) = \lambda_k$ for $i > 0$, where $\lambda_k$ is a constant, satisfying $0 < \lambda_k < 1$.

\smallskip

At time $t$, the stochastic process then proceeds as follows in order to obtain the configuration at time $t+1$, where $t \ge 0$.
\renewcommand{\labelenumi}{(\roman{enumi})}
\begin{enumerate}

\item For each $i$, $1 \le i \le t$, if $urn_i$ is non-empty then, with probability $\mu_k(i,t) = \lambda_k$, a mortality event occurs,
i.e. the ball in $urn_i$ is discarded.

\item Next, the ages of all balls remaining in the system are incremented by $1$, i.e. a ball in $urn_i$ is moved to $urn_{i+1}$, for each $i$.

\item Finally, with probability $b$, where $0 < b < 1$, a birth event occurs, i.e. a ball is inserted into $urn_0$.
\end{enumerate}
\smallskip

We observe that, at any time $t$, $urn_i$ is empty for $i > t$.
Note that we are assuming that the mortality rate is constant, i.e. $\mu_k(i,t) = \lambda_k$, which leads to the survival function of an  exponential distribution, as was shown in \cite{FENN15}. However, in principle, the mortality rate could be an arbitrary function, for example, power-law mortality leads to the survival function of a Weibull distribution. Note also that we are using a single birth probability $b$, rather than one for each process, since, as can be seen from (\ref{eq:expand}) below, $b$ is merely a scaling factor.
It follows that having different values for $b$ for the two processes is equivalent to adjusting the mixture probabilities.

\medskip

We now let $F_k(i,t) \ge 0$, with $k = 1$ or $2$, be a discrete function denoting the probability that there is a ball in $urn_i$ at time $t$, given that process $k$ is active. Initially, we set $F_k(0,0) = b$ and $F_k(i,0) = 0$ for all $i > 0$.

\smallskip

The dynamics of the model are captured by the following two equations, where we assume that $k=1$ or $2$ according to the outcome of the coin toss:
\begin{equation}\label{eq:init}
F_k(0,t) = b \ \ {\rm for} \ \ t \ge 0,
\end{equation}
and
\begin{equation}\label{eq:diff}
F_k(i+1,t+1) = F_k(i,t) - \mu_k(i,t) F_k(i,t) \ \ {\rm for} \ \ 0 \le i \le t.
\end{equation}
\smallskip

We can expand (\ref{eq:diff}) to obtain
\begin{equation}\label{eq:expand}
F_k(i+1,t+1) = b \prod_{j=1}^{i} \left( 1 - \mu_k(j,t-i+j) \right) = b (1 - \lambda_k)^i.
\end{equation}

Using the approximation $\ln(1-x) \approx -x$, which holds for small $x$, we see that the distribution is approximately exponential:
\begin{equation}\label{eq:approx}
F_k(i+1,t+1) \approx b \exp(- \lambda_k i).
\end{equation}

We note that, in (\ref{eq:approx}), we could use the more accurate second order approximation $\ln(1-x) \approx -x-x^2/2$.

\smallskip

We now define $F(i,t)$ analogously to $F_k(i,t)$ to take into account which of the two processes is active, i.e.
\begin{displaymath}
F(i,t) = p_1 F_1(i,t) + p_2 F_2(i,t).
\end{displaymath}

This yields the initial conditions $F(0,0) = b$, $F(i,0) = 0$ for all $i > 0$, and also $F(0,t) = b$ for $t \ge 0$.
Similarly, from (\ref{eq:expand}) and (\ref{eq:approx}) for $0 \le i \le t$, we obtain the exponential mixture:
\begin{align}\label{eq:mixture-t}
F(i+1,t+1) & = p_1 b (1 - \lambda_1)^i + p_2 b (1 - \lambda_2)^i  \\
& \approx p_1 b  \exp(- \lambda_1 i) + p_2 b  \exp(- \lambda_2 i). \nonumber
\end{align}

We note that, as mentioned in the beginning of this section, the model can be extended to more than two components; in this case
the sums in (\ref{eq:mixture-t}) would have the appropriate number of terms.

\section{Application of the Mixture Model}
\label{sec:app}


In the context of human dynamics, survival analysis has recently been applied to large data sets. These include the analysis of phone call durations \cite{MELO10}, the investigation of how long Wikipedia editors remain active \cite{ZHAN12}, and for predicting the likelihood of online content to become popular \cite{LEE12}.

\smallskip

The survival function $S(\theta)$ introduced in Section~\ref{sec:mixture} is usually estimated via a step function by computing the probability that an actor survives until time $\theta$, for $\theta = 1,2,\ldots, t$; this step function is known as the {\em Kaplan-Meier estimator} \cite{KAPL58,KLEI12}. By comparing (\ref{eq:expand}) with the Kaplan-Meier estimators for the survival functions \cite[equation~(2b)]{KAPL58} of the components of the mixture model, these are seen to be analogous to $F_k(i,t)$ for $k=1,2$, and an actor that was born at time $t-i$. We thus obtain
\begin{equation}\label{eq:survival-mixture}
S(i) \approx \frac{p_1 F_1(i,t) + p_2 F_2(i,t)}{b} = \frac{F(i,t)}{b}.
\end{equation}

In theory, the survival function $S(\theta)$ does not depend on the length $t$ of the trial. However, in practice, the Kaplan-Meier estimate will be more accurate for longer trials. Nevertheless, this estimate is more accurate when most of the actors are still present in the study, since the estimate may be inaccurate when there are only a few actors remaining \cite{RICH10}.

\medskip

As as a proof of concept for the exponential mixture model, we analysed the survival of queries in the top-10 Google Trends ``hot searches''  (\url{www.google.com/trends/hottrends}), that we used in \cite{FENN15}. Data was collected monthly for the top-10 ``hot searches'' over 114 months, from January 2004 until June 2013,
for six categories (together with their subcategories in each case): Business \& Politics (or simply Business), Entertainment, Nature \& Science (or simply Science), Shopping \& Fashion (or simply Shopping), Sports, and Travel \& Leisure (or simply Travel). The number of distinct queries per category over the period is shown in Table~\ref{table:google}. It is apparent from this statistic that the top-10 queries from Shopping change the least, while those from Entertainment change the most.

\begin{table}[ht]
\begin{center}
\begin{tabular}{|l|c|}\hline
Data set       & Number of queries \\ \hline \hline
Business       & 318 \\ \hline
Entertainment  & 1672 \\ \hline
Science        & 150 \\ \hline
Shopping       & 107 \\ \hline
Sports         & 774 \\ \hline
Travel         & 342 \\ \hline \hline
All Categories & 3363 \\ \hline
\end{tabular}
\end{center}
\caption{\label{table:google} Number of top-10 queries collected from Google Trends.}
\end{table}
\smallskip

In this data set, the balls are top-10 queries, a mortality event occurs when a query leaves the top-10 in a given month, and a birth event occurs when a new query joins the top-10 in a given month (note that time is discrete and is measured in months).
As we are considering an exponential mixture model with constant mortality for each component, queries leave the top-10 according to (\ref{eq:lambda-mix}).

\medskip

We next outline the methodology we have used to validate and evaluate the stochastic urn model presented in Section~\ref{sec:urn}. We then give further details, before discussing and analysing the results.

\renewcommand{\labelenumi}{(\Roman{enumi})}
\begin{enumerate}
\item First, to obtain estimates of $p_1$, $\lambda_1$ and $\lambda_2$, we performed nonlinear regression to the right-hand side of (\ref{eq:mixture-t}) using the Kaplan-Meier estimates computed from the raw data, for $i = 1,2,\ldots,114$.

\item We used the estimates of $p_1$, $\lambda_1$ and $\lambda_2$ from (I) to compute, for each $i$, the right-hand side of (\ref{eq:mixture-t}); we call this the {\em mixture data}. We then repeated the nonlinear regression using the mixture data as a quick ``sanity check'' that the new values of $p_1$, $\lambda_1$ and $\lambda_2$ obtained were consistent with those from (I).

\item Next we ran simulations using the parameters $p_1$, $\lambda_1$ and $\lambda_2$ from (I), and $b=0.9$. Using the averaged values of $F(i,t)$ from the simulations, we again repeated the nonlinear regression to obtain new values for $p_1$, $\lambda_1$ and $\lambda_2$; these were compared with those from (I) for consistency.

\item We computed the $D$ values for Kolmogorov-Smirnov tests to ascertain whether the Kaplan-Meier estimates, the mixture data and the averaged simulation data are likely to have all come from the same distribution.
\end{enumerate}
\smallskip

We first obtained the Kaplan-Meier estimates from the raw data sets for the six individual categories and for their aggregation (All Categories). Recalling that the survival function is approximated as in (\ref{eq:survival-mixture}), following (I), we used Matlab to obtain estimates for $p_1$, $\lambda_1$ and $\lambda_2$ by nonlinear least-squares regression of $S(i)$ on $i$ for fitting the right-hand side of (\ref{eq:mixture-t}), for $i = 1,2,\ldots,114$. The estimated parameters $p_1$, $\lambda_1$ and $\lambda_2$ are shown in the rows of Table~\ref{table:exp-km}, together with the coefficient of determination $R^2$ \cite{MOTU95}; the $R^2$ values show a very good fit for all of the categories. Nonlinear regression using the mixture data values computed from the right-hand side of (\ref{eq:mixture-t}), as in (II), yielded almost perfect fits, as expected.

\begin{table}[ht]
\begin{center}
\begin{tabular}{|l|c|c|c|c|}\hline
Data set       & $p_1$    & $\lambda_1$ & $\lambda_2$ & $R^2$  \\ \hline \hline
Business       & 0.4061 & 0.0136      & 0.4095      & 0.9893 \\ \hline
Entertainment  & 0.2013 & 0.0405      & 0.4793      & 0.9965 \\ \hline
Science        & 0.6693 & 0.0071      & 0.4391      & 0.9884 \\ \hline
Shopping       & 0.7403 & 0.0091      & 0.4880      & 0.9915 \\ \hline
Sports         & 0.2275 & 0.0215      & 0.4599      & 0.9829 \\ \hline
Travel         & 0.4483 & 0.0090      & 0.4364      & 0.9823 \\ \hline
All Categories & 0.2600 & 0.0154      & 0.4337      & 0.9894 \\ \hline
\end{tabular}
\end{center}
\caption{\label{table:exp-km} Nonlinear least-squares regression to a mixture of exponentials using the Kaplan-Meier estimates.}
\end{table}
\smallskip

To test the validity of the model, as in (III), we then carried out simulations in Matlab of the stochastic urn transfer model using the values of $p_1$, $\lambda_1$ and $\lambda_2$ from Table~\ref{table:exp-km}. We chose the value $b=0.9$ for all simulations, after running some sample simulations with other values of $b$. The value of $b$ is not critical since, as can be seen from (\ref{eq:mixture-t}), $b$ is merely a scaling factor. The simulations were run for $114$ steps, one for each month, for each of the categories, and this was repeated $10^6$ times. For each category, we then calculated the average value of $F(i,t)$ for $i=1,2,\ldots,t$, over the $10^6$ runs. Repeating the nonlinear regression using these average values gave the results shown in Table~\ref{table:exp-sim}. Comparing Table~\ref{table:exp-sim} with Table~\ref{table:exp-km} shows that all the values of $p_1$, $\lambda_1$ and $\lambda_2$ are very close.
Moreover, the very high values for $R^2$ confirm the validity of (\ref{eq:mixture-t}) as the solution to our model.

\begin{table}[ht]
\begin{center}
\begin{tabular}{|l|c|c|c|c|}\hline
Data set       & $p_1$    & $\lambda_1$ & $\lambda_2$ & $R^2$  \\ \hline \hline
Business       & 0.4052   & 0.0135      & 0.4037      & 0.9999 \\ \hline
Entertainment  & 0.1991   & 0.0402      & 0.4683      & 0.9996 \\ \hline
Science        & 0.6686   & 0.0071      & 0.4289      & 0.9994 \\ \hline
Shopping       & 0.7394   & 0.0091      & 0.4760      & 0.9996 \\ \hline
Sports         & 0.2269   & 0.0214      & 0.4504      & 0.9997 \\ \hline
Travel         & 0.4472   & 0.0089      & 0.4286      & 0.9997 \\ \hline
All Categories & 0.2598   & 0.0153      & 0.4260      & 0.9998 \\ \hline
\end{tabular}
\end{center}
\caption{\label{table:exp-sim} Nonlinear least-squares regression to a mixture of exponentials using the simulated data with $p$, $\lambda_1$ and $\lambda_2$ taken from Table~\ref{table:exp-km}.}
\end{table}
\smallskip

To compare the Kaplan-Meier estimates, the mixture data and the averaged simulation data, as in (IV), we performed three Kolmogorov-Smirnov 2-sample 2-tailed tests, as described in Section 6.6.4 of \cite{SIEG88}. Taking the null hypothesis to be that the Kaplan-Meier estimates, the mixture data and the averaged simulation data all come from the same population distribution, the critical value at significance level $\alpha=0.05$ is $0.1801$ for a sample of $114$ (number of months). The $D$ values for the three pairwise tests are shown in Table~\ref{table:ks-test}.
It can be seen that, in all cases, the values of the test statistic $D$ are well below the critical value. Hence, we cannot reject the null hypothesis at significance level $\alpha=0.05$. The values in the Sim-Mix column show that the distributions of the mixture data and averaged simulation data are extremely close, which is unsurprising since these are both derived directly from our model. We note that, even at significance level $\alpha=0.10$, where the critical value is $0.1616$, the null hypothesis cannot be rejected.

\begin{table}[ht]
\begin{center}
\begin{tabular}{|l|c|c|c|}\hline
Data set       & KM-Sim & KM-Mix & Sim-Mix \\ \hline \hline
Business       & 0.0761 & 0.0720 & 0.0126 \\ \hline
Entertainment  & 0.0640 & 0.0583 & 0.0247 \\ \hline
Science        & 0.0727 & 0.0598 & 0.0339 \\ \hline
Shopping       & 0.0351 & 0.0343 & 0.0289 \\ \hline
Sports         & 0.1015 & 0.0966 & 0.0226 \\ \hline
Travel         & 0.0841 & 0.0807 & 0.0170 \\ \hline
All Categories & 0.0856 & 0.0803 & 0.0175 \\ \hline
\end{tabular}
\end{center}
\caption{\label{table:ks-test} The $D$ values for the 2-sample 2-tailed Kolmogorov-Smirnov tests.}
\end{table}
\smallskip

In \cite{FENN15}, we used nonlinear least-squares regression to fit Weibull survival functions to the Kaplan-Meier estimates. The estimates obtained for the parameters $\gamma$ and $\rho$ are shown in Table~\ref{table:lsq-km-weibull}, taken from \cite{FENN15}.
We note that the shape parameters $\rho$ satisfy $-1 < \rho < 0$, which, as mentioned at the end of Section~\ref{sec:mixture}, implies that the Weibull survival functions can be expressed as mixtures of exponentials.
In order to compare our approximation by exponential mixtures with the Weibull survival functions obtained in \cite{FENN15},
we used least-squares curve fitting to fit Weibulls (\ref{eq:weibull}) to the exponential mixtures obtained from (\ref{eq:mixture-t}) and (\ref{eq:survival-mixture}), using the parameters from Table~\ref{table:exp-km}, in a similar manner to \cite{JIN10}. The estimated parameters $\gamma$ and $\rho$ are shown in Table~\ref{table:lsq-exp-weibull}, together with $R^2$. It can be seen that these estimates are very close to those in Table~\ref{table:lsq-km-weibull}.

\begin{table}[ht]
\begin{center}
\begin{tabular}{|l|c|c|c|}\hline
Data set       & $\gamma$ & $\rho$  & $R^2$\\ \hline \hline
Business       & 1.5576    & -0.8321 & 0.9818 \\ \hline
Entertainment  & 1.7877    & -0.7453 & 0.9957 \\ \hline
Science        & 0.7575    & -0.8318 & 0.9303\\ \hline
Shopping       & 0.4736    & -0.7608 & 0.9240 \\ \hline
Sports         & 3.1986    & -0.8666 & 0.9967 \\ \hline
Travel         & 3.5141    & -0.9291 & 0.9760 \\ \hline
All Categories & 4.5262    & -0.9134 & 0.9955 \\ \hline
\end{tabular}
\end{center}
\caption{\label{table:lsq-km-weibull} Nonlinear least-squares regression to Weibull survival functions using the Kaplan-Meier estimates.}
\end{table}
\smallskip

\begin{table}[ht]
\begin{center}
\begin{tabular}{|l|c|c|c|}\hline
Data set       & $\gamma$ & $\rho$  & $R^2$\\ \hline \hline
Business       & 1.5516 & -0.8315 & 0.9967 \\ \hline
Entertainment  & 1.7326 & -0.7389 & 0.9976 \\ \hline
Science        & 0.7500 & -0.8276 & 0.9962 \\ \hline
Shopping       & 0.4500 & -0.7510 & 0.9930 \\ \hline
Sports         & 3.0612 & -0.8613 & 0.9971 \\ \hline
Travel         & 3.2840 & -0.9250 & 0.9969 \\ \hline
All Categories & 4.3860 & -0.9108 & 0.9972 \\ \hline
\end{tabular}
\end{center}
\caption{\label{table:lsq-exp-weibull} Least-squares curve fitting of Weibulls to the exponential mixture survival functions.}
\end{table}
\smallskip

Finally, we note from Table~\ref{table:exp-km} that $\lambda_1$ is significantly smaller than $\lambda_2$.
It therefore follows, from the comment made after (\ref{eq:pi-mix}) regarding the decreasing failure rate, that
a substantial number of queries are popular for a short duration, because $\lambda_2$ is large, while the popularity of others lasts for much longer, because $\lambda_1$ is small. We note that a similar observation, in the context of the popularity of movies, was made in \cite{PAN10}. In particular, for the Google Trends data set, about 40\% of the queries survive for only a single month.

\section{Concluding Remarks}
\label{sec:conc}

In this paper, we propose a stochastic evolutionary urn-based model that can be applied to survival analysis scenarios in the context of human dynamics. In our model, actors survive in the system until they die and leave the system according to a constant mortality rate.
A solution to the equations describing the model was obtained in (\ref{eq:mixture-t}), and it was seen that this solution is approximately an exponential mixture. We then successfully applied our model to data on the survival of popular search engine queries. It is also potentially applicable to other data sets relating to human behaviour.

\smallskip

One important problem to be addressed is how to  choose the number of components in the mixture model. As discussed in
\cite[Chapter 6]{MCLA00}, several methods have been suggested to obtain a parsimonious model. Among these, a number of prominent methods are based on maximising a penalised log-likelihood function.

\smallskip

Generative models, like the one presented here, have the potential to explain observed social phenomena and, more specifically, social mechanisms and the emergence of collective behaviour, as discussed in the introduction. They allow us to gain insight into the underlying processes; in addition, they may be useful for more accurate prediction of, for example, online content, as discussed in \cite{TATA14}. With regards to our theme of human dynamics \cite{CAST09}, our stochastic mixture urn model contributes to a
better understanding of collective phenomena, such as popularity, and how such global behaviour emerges from the decisions of individuals from whom the statistics have been collected.
Another possible avenue for investigation is to consider other distribution mixtures, such as finite Weibull mixture models (see \cite{BUCA04}, where it is argued that the reliability of a system can be suitably modelled by a Weibull mixture model).

\section*{Acknowledgements}
We would like to thank Suneel Kingrani, who collected the Google Trends data and computed the Kaplan-Meier estimates for the data set.

\newcommand{\etalchar}[1]{$^{#1}$}

\end{document}